\documentclass[traditabstract,letter]{aa} 

\usepackage{verbatim}
\usepackage{graphicx}
\usepackage{natbib}
\bibpunct{(}{)}{;}{a}{}{,} 

\usepackage{txfonts}
\graphicspath{{./}{figs/}}

\begin{document}

\title{The broad iron K$\alpha$ line of Cygnus X-1 as seen by
  \textsl{XMM-Newton}  in the EPIC-pn modified timing mode }

   \author{Refiz Duro\inst{1} \and
          Thomas Dauser\inst{1} \and
          J\"orn Wilms\inst{1} \and
          Katja Pottschmidt\inst{2,3} \and
          Michael A. Nowak\inst{4}\and
          Sonja Fritz\inst{1,5} \and
          \mbox{Eckhard Kendziorra}\inst{5}\and
          Marcus G. F. Kirsch\inst{6}\and
          Christopher S. Reynolds\inst{7}\and
          R\"udiger Staubert\inst{5}
          }
  \institute{Dr.\ Karl Remeis-Sternwarte and Erlangen Centre
            for Astroparticle Physics,
            Friedrich-Alexander-Universit\"at Erlangen-N\"urnberg,
            Sternwartstra\ss{}e~7, 96049 Bamberg, Germany
            \and
            CRESST and NASA Goddard Space Flight Center, Astrophysics Science
            Division, Code 661, Greenbelt, MD 20771, USA
            \and 
            Center for Space Science and Technology, University of Maryland
            Baltimore County, 1000 Hilltop Circle, Baltimore, MD 21250, USA 
            \and
            Kavli Institute for Astrophysics, Massachusetts Institute
            of Technology, Cambridge, MA 02139, USA 
            \and
            Institut f\"ur Astronomie und Astrophysik, Eberhard Karls
            Universit\"at T\"ubingen, Sand 1, 72074 T\"ubingen, Germany
            \and
            European Space Agency, European Space Operations Centre,
            Robert-Bosch-Stra\ss{}e 5, 64293 Darmstadt, Germany
            \and
            Department of Astronomy and Maryland Astronomy Center for Theory and
            Computation, University of Maryland, College Park, MD 20742, USA 
          }

          \date{Received 9 June 2011 / accepted 21 July 2011}
 
          \abstract{We present the analysis of the broadened,
            flourescent iron K$\alpha$ line in simultaneous
            \textsl{XMM-Newton} and \textsl{RXTE} data from the black
            hole Cygnus X-1. The \textsl{XMM-Newton} data were taken
            in a modified version of the timing mode of the EPIC-pn
            camera. In this mode the lower energy threshold of the
            instrument is increased to 2.8\,keV to avoid telemetry
            drop outs due to the brightness of the source, while at
            the same time preserving the signal-to-noise ratio in the
            Fe K$\alpha$ band. We find that the best-fit spectrum
            consists of the sum of an exponentially cut off power-law
            and relativistically smeared, ionized reflection. The
            shape of the broadened Fe K$\alpha$ feature is due to
            strong Compton broadening combined with relativistic
            broadening. Assuming a standard, thin accretion disk, the
            black hole is close to rotating maximally.}

   \authorrunning {Duro et al.}
   \titlerunning{Broad Fe K$\alpha$ line of Cygnus X-1}
   \maketitle
%

\section{Introduction}
\label{sec:intro}

Relativistically broadened lines from accreting black holes have an
important diagnostic potential because they allow us to measure
directly properties of these black holes such as their spin, the
inclination of the accretion disk, and the emissivity profile of the
accretion disk \citep[e.g.,][]{Reynolds2003a}. While by now a large
number of high signal-to-noise ratio ($S/N$) measurements of the line
shape and parameters exist for many Galactic black holes
\citep{miller:07a} and for Active Galactic Nuclei
\citep[e.g.,][]{miniutti:07a,fabian:09a}, high $S/N$ measurements of
the line profile and black hole parameters of the canonical black
hole, Cygnus~X-1, are still lacking. This lack is mainly due to a
combination of source brightness, which causes the measurement to be
dominated by systematic effects in the detector calibration
\citep[see, e.g.,][]{miller:10a}, as well as a complex X-ray spectrum,
which necessitates broad band (3--100\,keV) measurements from multiple
satellites. See, however, \citet{Miller2002a} for recent work on the
Fe line region with \textsl{Chandra}'s gratings, and \citet[][and
references therein]{nowak:11a} on joint \textsl{Suzaku}/\textsl{RXTE}
measurements of the broad band spectrum.

In this \emph{letter} we present an analysis of the broad band
spectrum of \mbox{Cygnus X-1} using data from high $S/N$ observations
of the Fe line band from \textsl{XMM-Newton} and simultaneous,
3--120\,keV \textsl{RXTE} data. In Sect.~\ref{sec:mod_timing} we
present the modified timing mode, a special form of the
\textsl{XMM-Newton} timing mode, dedicated to bright sources with
fluxes up to a few 100\,mCrab. In Sect.~\ref{sec:observations} we
discuss the data reduction in greater detail.
Section~\ref{sec:relline} presents the analysis of the
relativistically broadened line and the reflection continuum.

\section{The EPIC-pn modified timing mode}\label{sec:mod_timing}
To determine the best Fe line parameters possible, very high $S/N$
observations are required in a time that is as short as possible to
avoid complications due to source spectral variability. The best
instrument currently fulfilling these requirements is the EPIC-pn
camera onboard \textsl{XMM-Newton} \citep{strueder:01a}. This camera
has two modes that provide the fast read-out capability required to
avoid pile-up in bright source data \citep{ness:10a}. Here, the term
``pile-up'' summarizes both ``energy pile-up'', where more than one
X-ray photon impacts the same CCD pixel during one CCD read-out cycle,
and ``pattern pile-up'' or ``grade migration'', where multiple X-ray
photons hit adjacent CCD pixels during one read-out cycle and the
resulting charge distribution in the CCD is wrongly interpreted as
originating from a single, higher energy, X-ray. For satellites with
moderate spatial resolution such as \textsl{XMM-Newton}, the latter
effect is more pronounced. In EPIC-pn's burst mode, fast exposures and
shifts are performed, interrupted by longer periods for data read out.
The livetime of this mode is only 3\%. In the timing mode, continuous
read out is available with a 99.5\% livetime. This mode is well suited
for observations with a low pile-up fraction for fluxes up to
$\sim$150\,mCrab, although the EPIC-pn telemetry allocation limits
this mode to observations of sources of fluxes below $\sim$100\,mCrab.
With $\sim$300\,mCrab, \mbox{Cyg~X-1} can therefore not be observed
with the standard timing mode.

In order to address the telemetry limitation we have proposed a
modification of the timing mode \citep{Kendziorra2004a}. In this
modified timing mode, the lower energy threshold for telemetered
events is increased from 0.15\,keV to 2.8\,keV, such that all events
in the Fe line band can be transferred without telemetry drop outs,
and the telemetry allocation to the EPIC-pn is maximized by switching
off the EPIC-MOS cameras, which are piled-up anyway for sources above
$\sim$35\,mCrab.

EPIC data from events in which the charge cloud produced by an X-ray
is distributed over more than one CCD pixel, so-called ``split
events'', are only recombined on ground. In the modified timing mode,
split partners with energies below 2.8\,keV are not telemetered,
which results in a slight degradation of the energy resolution and
necessitates the generation of a dedicated response matrix. As
discussed by \citet{Wilms2006a} and \citet{Fritz2008}, this effect can
be fully taken into account by measuring the energy dependent
probability distribution of split events from archival Timing Mode
observations and modifying the official Timing Mode response matrix
accordingly. A full description of the calibration will be given in a
forthcoming paper \citep{Duro2011b}. The \textsl{XMM-Newton} data
analysis and response matrix generation was based on the Science
Analysis Software (SAS), version 10.0.0, and the newest calibration
files.

It has recently been found that some Timing Mode observations are
affected by ``X-ray loading'' \citep{done:10a}, i.e., the
contamination of EPIC-pn's offset map by source X-rays. Since the
offset map is subtracted from the EPIC data before they are
telemetered to ground, loading distorts the X-ray spectrum and
effectively introduces a CCD-column dependent lower energy threshold.
In the modified timing mode, the offset map is determined with the
EPIC-pn filter in the blocking position. As no source X-rays enter the
offset map, the mode is unaffected by X-ray loading.

\begin{figure}
\centering
\includegraphics[width=\columnwidth]{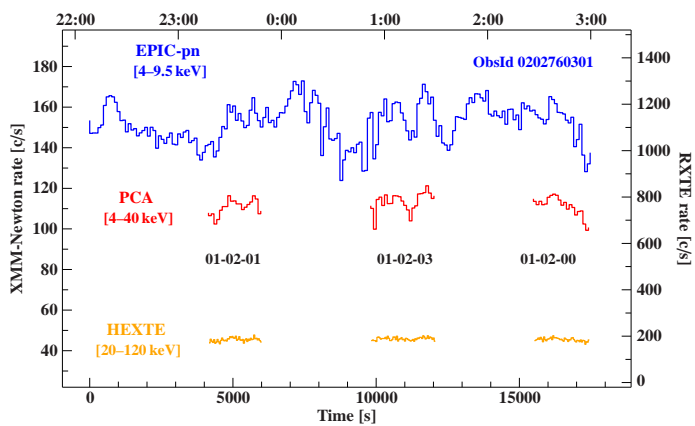}
\caption{\textsl{XMM-Newton} and \textsl{RXTE} lightcurve of
  Cygnus~X-1 on 2004 November 20/21. Here, and throughout the
  remainder of the letter, EPIC-pn data are shown in blue, PCA (PCU2)
  data in red, and HEXTE in orange, with resolutions of 100\,s, 96\,s,
  and 40\,s, respectively.}
\label{lightcurve}
\end{figure}

\section{Data reduction}\label{sec:observations}
\subsection{Data reduction}
In 2004 November and~December \textsl{XMM-Newton} performed four
modified timing mode observations of Cyg~X-1. Here, we concentrate on
the simultaneous \textsl{XMM-Newton} and \textsl{RXTE} observations
performed 2004 November 20 and~21 (\textsl{XMM-Newton} ID 0202760301),
during which Cyg~X-1 was in a transitional state between the hard
state and the thermally dominated soft state \citep{Fritz2008}. We
concentrate on this 17.4\,ks long observation since it shows the least
variability on timescales of minutes to hours (Fig.~\ref{lightcurve})
and thus provides the cleanest X-ray spectrum. Results from all
observations will be presented in a subsequent paper
\citep{Duro2011b}. The increased soft X-ray emission during the
intermediate state potentially leads to pile-up in the center of the
point spread function. For the present study, we therefore ignore the
innermost three columns. In order to avoid any transient effects due
to the 2.8\,keV lower energy threshold, we limit ourselves to the
4.0--9.5\,keV band.

The simultaneous 6.18\,ks of \textsl{RXTE} data (observation IDs
90104-01-02-00, -01, and -03) were reduced with \texttt{HEASOFT}
version 6.9. We used the 4--40\,keV data from the Proportional Counter
Array \citep[PCA;][]{jahoda:06a} and the 20--112\,keV data from the
High Energy X-ray Timing Experiment \citep[HEXTE;][]{rothschild:98a}
and ignored data taken within 10 minutes of passages through the South
Atlantic Anomaly and during times of high particle background. All
spectral analysis was done with the Interactive Spectral
Interpretation System \citep{Houck2000a,noble08}, HEXTE spectra were
rebinned to $S/N=30$, EPIC-pn to $S/N=100$, which resulted in a
resolution of 40\,eV at 6.4\,keV and oversampled the detector response
by a factor $\sim$4.
   
\begin{figure}
\centering
\includegraphics[width=\columnwidth]{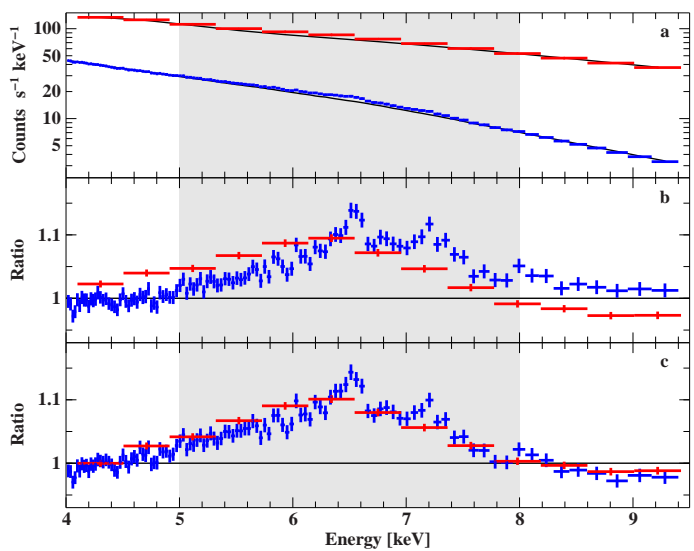}
\caption{\textbf{a} Fit of a simple power law to the 4.0--9.5\,keV
  data, ignoring the 5.0--8.0\,keV band (light gray area). \textbf{b}
  A broad excess due to the broad line, is visible. Note the energy shift
  between the PCA and the EPIC-pn. \textbf{c} Residuals after applying
  the gain shift correction.}\label{iron_line}
\end{figure}

\subsection{Simultaneous \textsl{XMM-Newton} and \textsl{RXTE} Fits}
The 3--120\,keV spectrum of Cyg~X-1 can be well described as the sum
of a soft excess with a temperature of $\lesssim$500\,eV, an
exponentially cutoff power law with a cutoff at a few 100\,keV, and a
Compton reflection component \citep[][and references
therein]{nowak:11a}. Alternatively, a variety of other, more physics
based, components have been used to describe the continuum
\citep{nowak:11a,malzac:08a,wilms06,markoff:05a}, but these are
virtually indistinguishable from this empirical model. In addition to
the continuum components, \textsl{Chandra}-HETGS observations show the
presence of a narrow line at 6.4\,keV emitted by neutral iron in
material surrounding the black hole at farther distances
\citep{hanke:09a}.

As shown in Fig.~\ref{iron_line}, a simple continuum model is not
sufficient to describe the data. Strong positive residuals in the
region between 5 and 7\,keV remain, which are due to the broad Fe
K$\alpha$ line. However, a slight mismatch in the \textsl{RXTE} and
\textsl{XMM-Newton} residuals is also apparent
(Figure~\ref{iron_line}b), with the EPIC line appearing to be shifted
to a slightly higher energy. This shift is confirmed by measuring the
energy of the 6.4\,keV narrow line feature, which is found at
6.6\,keV, inconsistent with all available \textsl{Chandra}
measurements for this feature in Cyg~X-1.

We attribute this difference to the treatment of charge transfer
efficiency (CTE) effects in the SAS. When observing bright sources
with a CCD such as the EPIC-pn, traps in the silicon crystal can be
saturated with source electrons, which improves the CTE. While the
latest SAS versions take this effect into account, their calibration
breaks down for very high count rates. An overcorrection of CTE
effects then leads to inferred photon energies that are too high. We
take this effect into account by correcting the assigned photon
energies with a model of the form
$E_\mathrm{real}=E_\mathrm{obs}/s+\Delta E$. The model parameters $s$
and $\Delta E$ are determined using measured energies of narrow
spectral components or from continuum fitting. As will be discussed in
greater detail by \citet{Duro2011b}, \textsl{Chandra}'s HETGS
observation (obsid 3814) was taken at the same flux level as our
\textsl{XMM-Newton}/\textsl{RXTE} observations. In addition to the
6.4\,keV line, \textsl{Chandra} reveals two Fe K$\alpha$ absorption
lines due to \ion{Fe}{xxv} and \ion{Fe}{xxvi} at 6.646\,keV and
6.955\,keV. Fits show that full consistency between the modified
timing mode spectrum and HETGS is obtained for $s=1.02$ and $\Delta
E=0$\,keV. This overcorrection of the EPIC energies by 2\% is in line
with typical CTE corrections for bright sources. An initial fit to the
simultaneous 4--10\,keV EPIC and PCA data agrees with this result
(Fig.~\ref{iron_line}c), showing that the CTE correction can also be
obtained from continuum fitting. The remaining differences between the
EPIC and PCA data are due to narrow features that are not resolved by
the PCA.

\section{The relativistic broadened Fe K$\alpha$ line of
  Cyg~X-1}\label{sec:relline} 
We now turn to modeling the broadened Fe line. To describe its profile we use
the \texttt{relline}-model of \citet{Dauser2010}, a relativistic line
model for thin disks around black holes with spins $-0.998\le a \le
+0.998$, where $a<0$ indicates that the angular momenta of the black
hole and the disk are antiparallel. We assume a radius-dependent line
emissivity per disk unit area that scales $\propto r^{-\epsilon}$,
where $\epsilon=3$ for a thin accretion disk. The disk extends from
the ($a$ dependent) innermost stable circular orbit to $400 GM/c^2$.

As expected, modeling the 4--10\,keV spectrum with a simple
relativistic line and a power-law continuum results in a bad fit
($\chi^2_\mathrm{red}\sim 1.5$) and unphysical line parameters as this
approach does not take reflection self-consistently into account. We
therefore extend the data to the full 4--120\,keV band provided by
EPIC-pn, PCA, and HEXTE, and model reflection of an exponentially
cutoff power law self-consistently using the reflection model
\texttt{reflionx} \citep{Ross2005a}. This model includes transitions
from the most important ions, including \ion{Fe}{vi} to
\ion{Fe}{xxvi}. In order to account for the relativistic smearing we
convolve the reflected spectrum with a relativistic convolution model
\citep[\texttt{relconv};][]{Dauser2010}. We assume that the angular
distribution of the fluorescent photons is isotropic and due to
photons emerging from a hot corona \citep{Svoboda2009a}. We also
included narrow K$\alpha$ absorption lines from \ion{Fe}{xxv} and
\ion{Fe}{xxvi} and a soft excess in the model.

\begin{figure}
\centering
\includegraphics[width=\columnwidth]{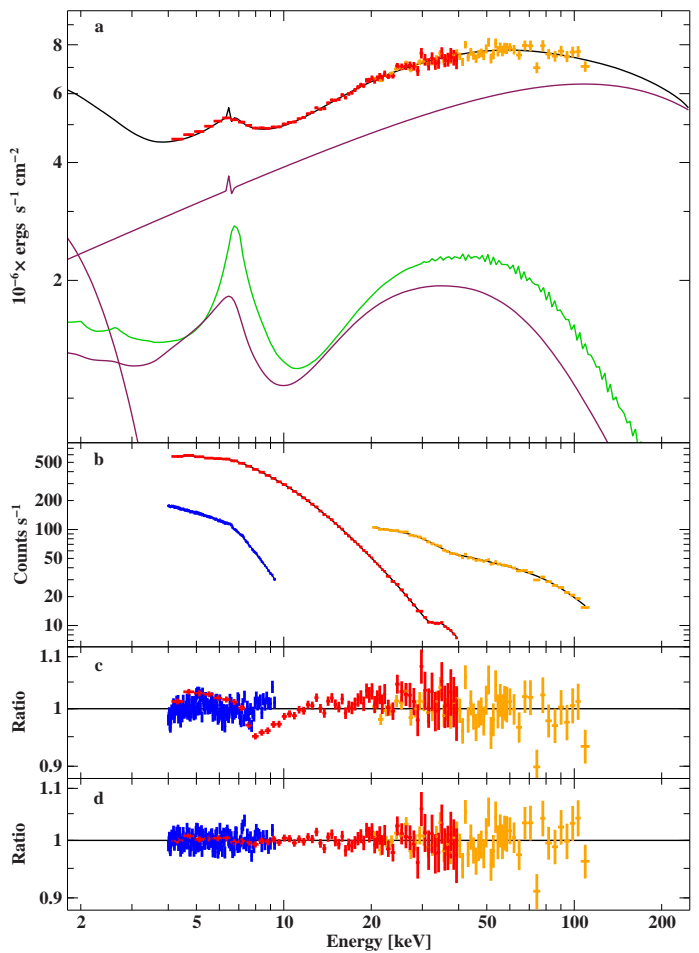}
\caption{\textbf{a} Unfolded XMM-Newton and RXTE data and best-fit
  model components. Green line: reflection continuum in the frame of
  the disk (note the strong Compton broadening; the ``jitter'' above
  50\,keV is due to the numerical resolution of the reflection model).
  Purple line: relativistically smeared reflection component.
  \textbf{b} Measured count rate spectra. \textbf{c} Residuals of the
  best-fit setting the relativistic convolution to zero. \textbf{d}
  Best-fit residuals including relativistic
  convolution.}\label{unfolded}
\end{figure}

\begin{table}
\centering
 \caption{Best fit results to gainshifted EPIC-pn, PCA, and
 HEXTE data assuming an exponentially cutoff power law, a narrow
  emission line at 6.4\,keV,  two resonant absorption lines, disk
  black body emission, and relativisticaly convolved
  reflection. Fluxes and normalizations, $A_i$, are determined
  relative to the  \textsl{RXTE}-PCA.  $c_\mathrm{EPIC}$ and
  $c_\mathrm{HEXTE}$ are flux normalization constants for
  the EPIC and HEXTE, respectively. Uncertainties are at the 90\% level
  for one  interesting parameter.} \label{table_fit}
\begin{tabular}{lcc}
 \hline
 Parameter & $\epsilon$ free & $\epsilon$ frozen \\
 \hline
 $\Gamma_\mathrm{pl}$  & $1.663^{+0.019}_{-0.017}$& $1.670\pm0.018$  \\
 $A_\mathrm{pl}$  & $1.17^{+0.09}_{-0.08}$ &  $1.23^{+0.07}_{-0.08}$\\ 
 $E_\mathrm{{fold}}$ [keV]  &  $290^{+80}_{-50}$ & $290^{+70}_{-50}$\\
 $A_\mathrm{bb}$  &  $\left(0.44^{+0.19}_{-0.07}\right)\times10^{4}$& $\left(0.54\pm0.07\right)\times10^{4}$\\
 \hline 
 $F_\mathrm{6.4\,keV} \mathrm{[cgs]}$ &  $\left(0.6\pm0.2\right)\times10^{-3}$ &  $\left(0.69^{+0.18}_{-0.20}\right)\times10^{-3}$\\
 $E_\mathrm{\ion{Fe}{xxv}\ K\alpha}$ [keV]  &    6.646 & 6.646 \\
 $\tau_1$  &   $\left(0.5\pm0.3\right)\times10^{-2}$ &  $\left(0.4\pm0.3\right)\times10^{-2}$\\
 $E_\mathrm{\ion{Fe}{xxvi}\ K\alpha}$ [keV]  &  6.955 & 6.955 \\
 $\tau_\mathrm{{2}}$  & $\le2\times10^{-3}$ &$\le1.9\times10^{-3}$\\
 \hline
 $\xi$  $[\mathrm{erg}\,\mathrm{cm}\,\mathrm{s}^{-1}]$  &  $1700^{+300}_{-400}$  &  $1400^{+300}_{-200}$\\ 
 $A_\mathrm{reflionx}$  & $\left(1.2\pm0.3\right)\times10^{-5}$ & $\left(1.4\pm0.3\right)\times10^{-5}$\\
 $\mathrm{Fe}/\mathrm{Fe}_\odot$  & $1.6^{+0.5}_{-0.4}$  &  $1.7^{+0.5}_{-0.4}$\\ 
 \hline
 $a$  & $-0.1\pm0.4$  &$0.88^{+0.07}_{-0.11}$\\
 $i [\mathrm{deg}]$ & $36^{+2}_{-4}$  &   $32\pm2$ \\
 $\epsilon$  &   $10^{+0}_{-6}$ & 3 \\ 
 \hline
 $c_\mathrm{HEXTE}$  &    $0.830\pm0.005$ &  $0.830^{+0.006}_{-0.005}$ \\
 $c_\mathrm{EPIC}$  &  $0.785\pm0.004$ & $0.785\pm0.004$ \\
 $s_\mathrm{gainshift}$  &    $1.0230^{+0.0019}_{-0.0017}$ &$1.0240^{+0.0019}_{-0.0018}$  \\
 \hline
 $\chi^2$/dof & 254/237 &  261/238 \\
 $\mathrm{\chi^{2}_{red}}$ &  1.08 & 1.10 \\
 \hline
 \hline
\end{tabular}
\end{table}

\begin{figure}
\centering
\includegraphics[width=\columnwidth]{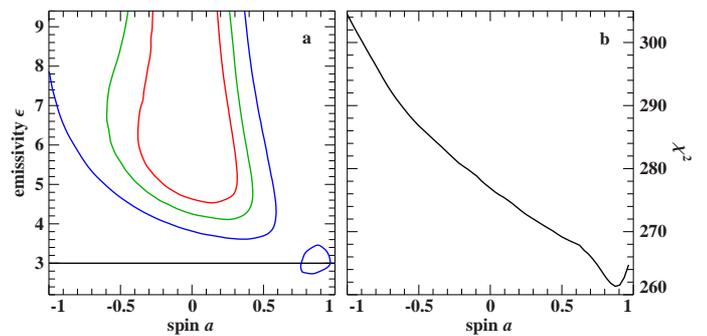}
\caption{\textbf{a} $\chi^2$ significance contours for the emissivity
  index, $\epsilon$, and the spin, $a$. Contours are based on $\Delta
  \chi^2=2.30$, 4.61, and 9.21, i.e., 68\%, 90\%, and 99\% confidence
  for two parameters of interest, with respect to the best fit case
  with $\epsilon$ let free. Two minima are apparent, one indicating a
  low-spin black hole at an unphysically large $\epsilon$, and one
  indicating an almost maximally spinning black hole consistent with
  emission from a thin accretion disk ($\epsilon=3$). \textbf{b}
  $\chi^2$ as a function of $a$ for $\epsilon=3$. }\label{contourplot}
\end{figure}

We first modeled the continuum using the canonical value of $\epsilon=3$
for a thin accretion disk. As shown in Fig.~\ref{unfolded} and
Table~\ref{table_fit}, this model describes the 4--120\,keV energy
spectrum well, with a reduced $\chi^2_\mathrm{red}=1.10$. The line
profile indicates a high angular momentum of the black hole
($a=0.88^{+0.08}_{-0.10}$, Fig.~\ref{contourplot}b), with additional line
broadening caused by  strong Compton broadening in the strongly
ionized reflector (ionization parameter
$\xi=1400^{+300}_{-100}\,\mathrm{erg}\,\mathrm{cm}\,\mathrm{s}^{-1}$),
which smears out all other discrete features such as edges in the
reflection spectrum (Fig.~\ref{unfolded}, green line). The inferred
inclination is $32^\circ\pm 2^\circ$, which agrees well with most
estimates for the inclination of the system (e.g., $32^\circ \le i \le
40^\circ$, \citealt{Ninkov1987a}, $i<55^\circ$, \citealt{Sowers1998a},
or $36^\circ\le i\le 67^\circ$ \citealt{Davis1983a}. The Fe abundance
is slightly higher than solar.

In order to check the uniqueness of this thin disk solution, we
performed where $\epsilon$ was let free. Figure~\ref{contourplot}
shows significance contours for this case. Most best-fit parameters
are consistent between both fits, specifically, the inclination
remains around $35^\circ$ and the reflector remains strongly ionized.
The emissivity, however, increases to $\epsilon=10$ (the hard limit)
while $a$ decreases to 0. Broad lines can therefore be generated
either by strongly concentrating all available emissivity to the
innermost regions of a disk around a Schwarzschild black hole, or by
emission from a standard disk around a maximally rotating black hole.
Both solutions are comparable in statistical quality\footnote{For
  $\Delta\chi^2=7$ and 235, respectively 236 degrees of freedom, the
  $F$-test indicates no improvement by introducing a free disk
  emissivity at the 99\% level.}. Given that most physical scenarios
for very steep emissivity profiles such as strongly torqued accretion
disks \citep{agol:00a} or strong illumination of the innermost disk
due to strong light bending from a high latitude source (``lamppost
models''; \citealt{martocchia:96a}) also require high $a$, the high
spin solution is preferred on physical grounds. Cygnus~X-1 is
therefore plausibly close to rotating maximally.

To conclude, this first analysis of a data set taken in the EPIC-pn
modified timing mode confirms the capability of the EPIC-pn camera to
observe sources up to several 100\,mCrab, yielding high $S/N$
measurements of the Fe line band. Coupled with broad band data from
other satellites such as \textsl{RXTE}, \textsl{INTEGRAL}, or
\textsl{Suzaku}, the modified timing mode allows the high precision
measurement of the physical parameters of black hole candidates,
including their inclination, spin, and emissivity profile. Applying
this analysis to Cyg~X-1 shows that its black hole is likely a Kerr
black hole if one assumes the accretion disk to be flat and
illuminated from a Compton corona. This result is consistent with spin
measurements from the accretion disk continuum \citep{gou:11a}. Being
in a high mass X-ray binary, the black hole in Cyg~X-1 is
comparatively young. Our result indicates that it must have been born
at close to maximum spin as expected in typical scenarios for black
hole formation in supernovae \citep{macfadyen:99a}. In a next step,
all four modified timing mode observations will be considered
\citep{Duro2011b}.

\begin{acknowledgements}
  This work was partly supported by the European Commission under
  contract ITN215212 ``Black Hole Universe'' and by the
  Bundesministerium f\"ur Wirtschaft and Technologie under Deutsches
  Zentrum f\"ur Luft- und Raumfahrt grants 50\,OR\,0701 and
  50\,OR\,1001. This paper is based on observations obtained with
  \textsl{XMM-Newton}, an ESA science mission with instruments and
  contributions directly funded by ESA member states and NASA. We
  thank Norbert Schartel and the \textsl{XMM-Newton} operations team
  for agreeing to perform observations in a new and untested mode,
  Maria D\'iaz-Trigo for many useful discussions on CTE and pile-up
  effects in the EPIC-pn camera, Manfred Hanke for his significant
  input concerning the data analysis and interpretation for black hole
  X-ray data, and J\'erome Rodriguez for his many insightful comments.
\end{acknowledgements}


\begin{thebibliography}{32}
\expandafter\ifx\csname natexlab\endcsname\relax\def\natexlab#1{#1}\fi

\bibitem[{{Agol} \& {Krolik}(2000)}]{agol:00a}
{Agol}, E. \& {Krolik}, J.~H. 2000, \apj, 528, 161

\bibitem[{{Dauser} {et~al.}(2010){Dauser}, {Wilms}, {Reynolds}, \&
  {Brenneman}}]{Dauser2010}
{Dauser}, T., {Wilms}, J., {Reynolds}, C.~S., \& {Brenneman}, L.~W. 2010,
  \mnras, 409, 1534

\bibitem[{{Davis} \& {Hartmann}(1983)}]{Davis1983a}
{Davis}, R. \& {Hartmann}, L. 1983, \apj, 270, 671

\bibitem[{Done \& {D\'iaz Trigo}(2009)}]{done:10a}
Done, C. \& {D\'iaz Trigo}, M. 2009, \mnras, 407, 2287

\bibitem[{Duro {et~al.}(2011)Duro, Dauser, Wilms, Fritz, Pottschmidt, Nowak,
  Kendziorra, Kirsch, Reynolds, \& Staubert}]{Duro2011b}
Duro, R., Dauser, T., Wilms, J., {et~al.} 2011, \aap, in prep.

\bibitem[{{Fabian} {et~al.}(2009){Fabian}, {Zoghbi}, {Ross}, {Uttley}, {Gallo},
  {Brandt}, {Blustin}, {Boller}, {Caballero-Garcia}, {Larsson}, {Miller},
  {Miniutti}, {Ponti}, {Reis}, {Reynolds}, {Tanaka}, \& {Young}}]{fabian:09a}
{Fabian}, A.~C., {Zoghbi}, A., {Ross}, R.~R., {et~al.} 2009, \nat, 459, 540

\bibitem[{Fritz(2008)}]{Fritz2008}
Fritz, S. 2008, Dissertation, Eberhard-Karls Universit\"at T\"ubingen

\bibitem[{{Gou} {et~al.}(2011){Gou}, {McClintock}, {Reid}, {Orosz}, {Steiner},
  {Narayan}, {Xiang}, {Remillard}, {Arnaud}, \& {Davis}}]{gou:11a}
{Gou}, L., {McClintock}, J.~E., {Reid}, M.~J., {et~al.} 2011, \apj, submitted
  (arXiv:1106.3690)

\bibitem[{{Hanke} {et~al.}(2009){Hanke}, {Wilms}, {Nowak}, {Pottschmidt},
  {Schulz}, \& {Lee}}]{hanke:09a}
{Hanke}, M., {Wilms}, J., {Nowak}, M.~A., {et~al.} 2009, \apj, 690, 330

\bibitem[{{Houck} \& {Denicola}(2000)}]{Houck2000a}
{Houck}, J.~C. \& {Denicola}, L.~A. 2000, in ASP Conf.\ Ser. 216, 
  Astronomical Data Analysis Software and Systems IX, ed. N.~Manset,
  C.~Veillet, \& D.~Crabtree, 591--595

\bibitem[{Jahoda {et~al.}(2006)Jahoda, Markwardt, Radeva, Rots, Stark, Swank,
  Strohmayer, \& Zhang}]{jahoda:06a}
Jahoda, K., Markwardt, C.~B., Radeva, Y., {et~al.} 2006, \apjs, 163, 401

\bibitem[{{Kendziorra} {et~al.}(2004){Kendziorra}, {Wilms}, {Haberl}, {Kirsch},
  {Martin}, \& {Nowak}}]{Kendziorra2004a}
{Kendziorra}, E., {Wilms}, J., {Haberl}, F., {et~al.} 2004, in {UV} to
  Gamma-Ray Space Telescope Systems, ed. G.~Hasinger \& M.~J.~L. Turner, Proc.\
  SPIE 5488 (Bellingham, WA: SPIE), 613--622

\bibitem[{{MacFadyen} \& {Woosley}(1999)}]{macfadyen:99a}
{MacFadyen}, A.~I. \& {Woosley}, S.~E. 1999, \apj, 524, 262

\bibitem[{{Malzac} {et~al.}(2008){Malzac}, {Lubi{\'n}ski}, {Zdziarski},
  {Cadolle Bel}, {T{\"u}rler}, \& {Laurent}}]{malzac:08a}
{Malzac}, J., {Lubi{\'n}ski}, P., {Zdziarski}, A.~A., {et~al.} 2008, \aa, 492,
  527

\bibitem[{Markoff {et~al.}(2005)Markoff, Nowak, \& Wilms}]{markoff:05a}
Markoff, S., Nowak, M.~A., \& Wilms, J. 2005, \apj, 635, 1203

\bibitem[{{Martocchia} \& {Matt}(1996)}]{martocchia:96a}
{Martocchia}, A. \& {Matt}, G. 1996, \mnras, 282, L53

\bibitem[{Miller(2007)}]{miller:07a}
Miller, J. 2007, \araa, 45, 441

\bibitem[{{Miller} {et~al.}(2010){Miller}, {D'A{\`i}}, {Bautz},
  {Bhattacharyya}, {Burrows}, {Cackett}, {Fabian}, {Freyberg}, {Haberl},
  {Kennea}, {Nowak}, {Reis}, {Strohmayer}, \& {Tsujimoto}}]{miller:10a}
{Miller}, J.~M., {D'A{\`i}}, A., {Bautz}, M.~W., {et~al.} 2010, \apj, 724, 1441

\bibitem[{{Miller} {et~al.}(2002){Miller}, {Fabian}, {Wijnands}, {Remillard},
  {Wojdowski}, {Schulz}, {Di Matteo}, {Marshall}, {Canizares}, {Pooley}, \&
  {Lewin}}]{Miller2002a}
{Miller}, J.~M., {Fabian}, A.~C., {Wijnands}, R., {et~al.} 2002, \apj, 578, 348

\bibitem[{{Miniutti} {et~al.}(2007){Miniutti}, {Fabian}, {Anabuki}, {Crummy},
  {Fukazawa}, {Gallo}, {Haba}, {Hayashida}, {Holt}, {Kunieda}, {Larsson},
  {Markowitz}, {Matsumoto}, {Ohno}, {Reeves}, {Takahashi}, {Tanaka},
  {Terashima}, {Torii}, {Ueda}, {Ushio}, {Watanabe}, {Yamauchi}, \&
  {Yaqoob}}]{miniutti:07a}
{Miniutti}, G., {Fabian}, A.~C., {Anabuki}, N., {et~al.} 2007, \pasj, 59, 315

\bibitem[{Ness {et~al.}(2010)}]{ness:10a}
Ness, J.~U. {et~al.} 2010, XMM-Newton Users Handbook, issue 2.8.1, ESA: ESAC,
  Villafranca

\bibitem[{{Ninkov} {et~al.}(1987){Ninkov}, {Walker}, \& {Yang}}]{Ninkov1987a}
{Ninkov}, Z., {Walker}, G.~A.~H., \& {Yang}, S. 1987, \apj, 321, 425

\bibitem[{Noble \& Nowak(2008)}]{noble08}
Noble, M.~S. \& Nowak, M.~A. 2008, \pasp, 120, 821

\bibitem[{Nowak {et~al.}(2011)Nowak, Hanke, Trowbridge, Markoff, Wilms,
  Pottschmidt, \& Coppi}]{nowak:11a}
Nowak, M.~A., Hanke, M., Trowbridge, S.~N., {et~al.} 2011, \apj, 728, 13

\bibitem[{{Reynolds} \& {Nowak}(2003)}]{Reynolds2003a}
{Reynolds}, C.~S. \& {Nowak}, M.~A. 2003, \physrep, 377, 389

\bibitem[{{Ross} \& {Fabian}(2005)}]{Ross2005a}
{Ross}, R.~R. \& {Fabian}, A.~C. 2005, \mnras, 358, 211

\bibitem[{Rothschild {et~al.}(1998)Rothschild, Blanco, Gruber, Heindl,
  {MacDonald}, Marsden, Pelling, Wayne, \& Hink}]{rothschild:98a}
Rothschild, R.~E., Blanco, P.~R., Gruber, D.~E., {et~al.} 1998, \apj, 496, 538

\bibitem[{{Sowers} {et~al.}(1998){Sowers}, {Gies}, {Bagnuolo}, {Shafter},
  {Wiemker}, \& {Wiggs}}]{Sowers1998a}
{Sowers}, J.~W., {Gies}, D.~R., {Bagnuolo}, W.~G., {et~al.} 1998, \apj, 506,
  424

\bibitem[{{Str{\"u}der} {et~al.}(2001){Str{\"u}der}, {Briel}, {Dennerl},
  {Hartmann}, {Kendziorra}, {Meidinger}, {Pfeffermann}, {Reppin}, {Aschenbach},
  {Bornemann}, {Br{\"a}uninger}, {Burkert}, {Elender}, {Freyberg}, {Haberl},
  {Hartner}, {Heuschmann}, {Hippmann}, {Kastelic}, {Kemmer}, {Kettenring},
  {Kink}, {Krause}, {M{\"u}ller}, {Oppitz}, {Pietsch}, {Popp}, {Predehl},
  {Read}, {Stephan}, {St{\"o}tter}, {Tr{\"u}mper}, {Holl}, {Kemmer}, {Soltau},
  {St{\"o}tter}, {Weber}, {Weichert}, {von Zanthier}, {Carathanassis}, {Lutz},
  {Richter}, {Solc}, {B{\"o}ttcher}, {Kuster}, {Staubert}, {Abbey}, {Holland},
  {Turner}, {Balasini}, {Bignami}, {La Palombara}, {Villa}, {Buttler},
  {Gianini}, {Lain{\'e}}, {Lumb}, \& {Dhez}}]{strueder:01a}
{Str{\"u}der}, L., {Briel}, U., {Dennerl}, K., {et~al.} 2001, \aap, 365, L18

\bibitem[{{Svoboda} {et~al.}(2009){Svoboda}, {Dov{\v c}iak}, {Goosmann}, \&
  {Karas}}]{Svoboda2009a}
{Svoboda}, J., {Dov{\v c}iak}, M., {Goosmann}, R., \& {Karas}, V. 2009, \aap,
  507, 1

\bibitem[{Wilms {et~al.}(2006{\natexlab{a}})Wilms, Kendziorra, Nowak,
  Pottschmidt, Haberl, Kirsch, \& Fritz}]{Wilms2006a}
Wilms, J., Kendziorra, E., Nowak, M.~A., {et~al.} 2006{\natexlab{a}}, in Proc.\
  X-ray Universe 2005, ed. A.~Wilson, ESA SP-604 (Noordwijk: ESA Publications
  Division), 217--222

\bibitem[{Wilms {et~al.}(2006{\natexlab{b}})Wilms, Nowak, Pottschmidt,
  {et~al.}}]{wilms06}
Wilms, J., Nowak, M.~A., Pottschmidt, K., {et~al.} 2006{\natexlab{b}}, \aap,
  447, 245

\end{thebibliography}
\end{document}